# The open access effect in social media exposure of scholarly articles: A matched-pair analysis

Huixu Li[1], Lanjian Liu[1], Xianwen Wang[2]*

1. Chang'An University, Xi'an, China
2. WISE Lab, Institute of Science of Science and S&T Management, Dalian University of Technology, Dalian, China
* Corresponding author, xianwenwang@dlut.edu.cn

**Abstract**
Scholarly journals are increasingly using social media to share their latest research publications and communicate with their readers. Having a presence on social media gives journals a platform to raise their profile and promote their content. This study compares the number of clicks received when journals provide two types of links to subscription articles: open access (OA) and paid content links. We examine the OA effect using unique matched-pair data for the journal *Nature Materials*. Our study finds that OA links perform better than paid content links. In particular, when the journal does not indicate that a link to an article is an OA link, there is an obvious drop in performance against clicks on links indicating OA status. OA has a positive effect on the number of clicks in all countries, but its positive impact is slightly greater in developed countries. The results suggest that free content is more attractive to users than paid content. Social media exposure of scholarly articles promotes the use of research outputs. Combining social media dissemination with OA appears to enhance the reach of scientific information. However, extensive further efforts are needed to remove barriers to OA.

## 1. Introduction

Social media has become an important platform for disseminating scientific information, enabling the public to both obtain and discuss research findings. The ease of employing social media channels to disseminate information and communicate makes their use attractive to researchers. Many scientists use social media to follow new discoveries and engage in scientific discussion. Scientists, editors, and publishers are increasingly focusing on expanding their own influence and/or that of their research results through social media exposure. Among the main social media platforms, Twitter and Facebook have emerged as the key outlets for disseminating scientific information (Lévy, 2018).

*1.1 Social media and altmetrics*

Social media offers a valuable platform for connecting scientists with society in general. In a 2015 survey of members of the American Association for the Advancement of Science (AAAS), 71% of those surveyed reported that their specialty area is either of some or of considerable interest to the public (Rainie et al., 2015). Moreover, the majority of tweets about scientific research seem to emanate from the general public, as suggested by the finding by Tsou et al. (2015) that only 34.4% of individual Twitter users who tweet links to academic articles hold a Ph.D.

With the constant increase in scholarly communication on the Internet, especially on social media platforms, traditional citation-based metrics are no longer sufficient to measure the influence of research outputs; there is a growing need to incorporate online interactions concerning scholarly papers into the impact analysis framework so as to reflect the broad and rapid impact of scholarship in this burgeoning ecosystem (Priem et al., 2010).

Social media exposure can increase article visits, with previous studies showing a strong and significant correlation between altmetrics and article visits (Wang et al., 2016, 2017). Although it is difficult to track whether someone tweeting about or viewing a paper via social media ultimately cites it, it is undeniable that social media engagement, including retweets, likes, and Facebook reactions, boosts the societal impact of research, regardless of whether the engagement is by scientists or members of the general public (Cui et al., 2018). Therefore, although the relationship between altmetrics and citation indicators remains vague, "altmetrics may capture a different aspect of societal impact to that seen by reviewers" (Bornmann, 2019). Social media metrics are often seen as positive indicators of public interest in science (Haustein et al., 2015); however, there are many factors affecting social media use that can complicate the results, e.g. differences in scholarly communication styles depending on age, academic rank, gender, discipline, country, and language (Sugimoto et al., 2017).

*1.2 Social media use by journals*

Users are kept informed of new resources on specified topics by SDI (selective dissemination of information) such as the current-awareness services informing library users about new acquisitions. Publishers often have a page listing RSS feeds or email alerts for all their journals. However, traditional academic communication still mainly relies on end users searching directly for the relevant knowledge they need from a library, whereas social media "pushes" knowledge directly to end users through websites and platforms such as Facebook and Twitter (Allen et al., 2013). Often, content promoted through social media is not closely related to end users' current research topics, which makes it like a kind of information encountering. When a user notices a paper pushed by journals through a social media post that interests them, they may then click on the link to the paper embedded in the post and ultimately cite it in their own research if it is sufficiently relevant: there are, thus, many steps between attracting social media attention and being cited (Wang et al., 2016).

The benefits for journals of participating in social media include self-promotion, improvement of journal metrics, and increasing knowledge translation (Hawkins et al., 2014, 2017; O'Kelly et al., 2017; Trueger, 2018). In order to promote research results, improve outreach, and connect with a potentially wider audience, including the non-academic community, journals are increasingly adopting social media tools and evaluating their impact on the scientific community (Sugimoto et al., 2017; Zedda & Barbaro, 2015), among which the Nature Publishing Group (NPG) is the most active publisher on social media (Zedda & Barbaro, 2015). Journals participate in social media in a variety of ways. Usually, publishers and journals use social media accounts to share scientific news and recently published research. Others have taken more aggressive steps, including using blogs, podcasts, and more interactive methods such as online journal clubs (Hawkins et al., 2017; Trueger, 2018). Some publishers and journals also encourage authors to promote their own work via social media platforms by providing tweetable abstracts, thus generating greater publicity for the journals they publish (Darling, Shiffman, Côté, & Drew, 2013; Kelly & Delasalle, 2012; Sugimoto et al., 2017). However, shared links to scholarly papers behind paywalls may not be accessible, which reduces the value of social media exposure of scholarly papers.

*1.3 Open Access*

Open access (OA) removes major obstacles to making research outputs freely available online and thus accessible to all members of society (Wang et al., 2015, 2018). OA improves access to research for all. In recent years, it has gradually become another important factor affecting the diffusion of academic articles (Ezema et al., 2017). The OA advantage has been confirmed in various empirical studies. Compared to non-OA articles, OA articles are cited more often, ultimately generating more recognition for their authors (Koler et al., 2014; Sotudeh et al., 2018; Wang et al., 2015) — this is referred to as the OA effect.

There are two main types of OA: Gold OA and Green OA (Björk et al., 2014; Harnad et al., 2008). The former type of publication is available directly from the publishers for free; the latter is self-archived by the authors in an OA repository (Laakso et al., 2011; Tennant et al., 2016). In recent years, a new kind of OA initiative has emerged: for example, SharedIt, developed by Springer Nature, allows authors and subscribers to share a link to a read-only full-text version of an article. Shareable links can be posted anywhere, including on social media, institutional repositories, and authors' own websites. All readers can download, print, save as a PDF, or view the full-text HTML version of OA articles shared using SharedIt; for subscription articles, all readers can view the full-text HTML version, but only subscribers can download, print, or save as a PDF. The use of SharedIt by journals such as *Nature Materials* to share their latest published research on social media may be considered a new attempt to combine OA and social media.

Access to information is crucial to supporting development, and open access to scientific information is one solution to the needs of developing countries (Schöpfel, 2017). OA is a matter of special concern in these countries, which have less money available to fund or publish research and less to access papers behind paywalls. For researchers in developing countries, OA enables their own research more visible to others and makes research more accessible for them. OA helps to integrate the work of scientists everywhere into the global knowledge base, reduce the isolation of researchers, and improve opportunities for funding and international collaboration (Suber & Arunachalam, 2005). However, there are many barriers restricting or preventing access to scholarly resources, including copyright laws that restrict access, subscription policies, and embargo periods on scholarly papers set by publishers. Some journals, like NEJM and PNAS, provide delayed free online access to their research articles (six months after publication) for readers from developing countries. A delayed OA policy may mitigate the access crisis but does not solve it; immediate open access is more important for developing countries. However, immediate open access to subscription articles would affect commercial interests and be unacceptable to publishers.

In previous studies, the OA advantage was confirmed by comparing the bibliometric performance of OA articles and non-OA articles; however, there is some controversy about the OA advantage. Confounding factors (the author self-selection bias in making higher-quality studies available through OA; the absence of an appropriate control group of non-OA articles with which to compare the bibliometric figures; the conflation of pre-publication vs. published/publisher versions of articles, etc.) make demonstrating a real bibliometric difference difficult (Ottaviani, 2016). There are several unanswered question. Do OA links (links to OA articles) get more reactions than paid content links (links to subscription articles) on social media? What is the difference in the OA effect between developed and developing countries? Exploring these questions would help us to measure the OA advantage of research papers with social media exposure. The present study differs from previous ones in that it explores the OA effect in the social media context by employing a unique matched-pair dataset of subscription articles with both OA (read-only) and paid (without restriction) versions. The main research questions in this study are:

RQ1. For the same subscription article, does the OA link receive more clicks than the paid content link, and, if so, to what degree?

RQ2. For the same subscription article, does the OA link have visitors from a wider range of regions than the paid content link? Does OA benefit people from developing countries more than those in developed countries?

RQ3. What is the relationship between social media engagement (retweets, likes, and Facebook reactions) and clicks on links? Are the effects of social media engagement on OA links different from those on paid content links?

## 2. Materials
*2.1 Data description*
Data for this study were retrieved from three platforms: Twitter, Facebook, and Bitly. We selected the journal *Nature Materials* as our research object because it has unique data suitable for matched-pair analysis of the OA effect. The editors of *Nature Materials* share the journal's latest published research on Twitter and Facebook through the official account @NatureMaterials. The aim is to share "cutting-edge research in materials science, also at the interface with physics, chemistry, biology, and medicine" (https://twitter.com/NatureMaterials). As of August 2019, the Twitter account had 31,700 followers and the Facebook page had 3,080 followers. Between May 2017 and September 2018, the editors provided two shortlinks to the same paper in a tweet or post: one OA link and one paid content link (see the example in Figure 1). The OA link provided access to view-only versions of subscription articles, meaning that everyone was able to read the subscription article through the link but not to save, download, or print it, whereas the paid content link was a traditional link through which only subscribers were able to view, download, print, and save the full-text paper. Unique data on the number of clicks for each link to the same paper enable easy comparison of the OA advantage. After September 2018, @NatureMaterials adjusted its posting style by only providing an OA shortlink (e.g., the link https://go.nature.com/2wD2NFW shown in Figure 1). However, because it does not clearly indicate the link's OA status, the audience cannot know whether the link is OA. The adjustment thus allows us to examine the influence or lack of influence of indicating OA status on how many times links are clicked on. In other words, would links that were, in fact, OA but did not carry the OA sign receive fewer clicks than those with the OA sign?

Springer Nature journals use Bitly to shorten, create, and share links for URLs. Shortlinks beginning with "go.nature.com" are actually Springer Nature's custom-branded links supported by Bitly. We used the Bitly API to track each shortlink's performance, including clicks, referrers, domains, and locations (countries) of the clickers.

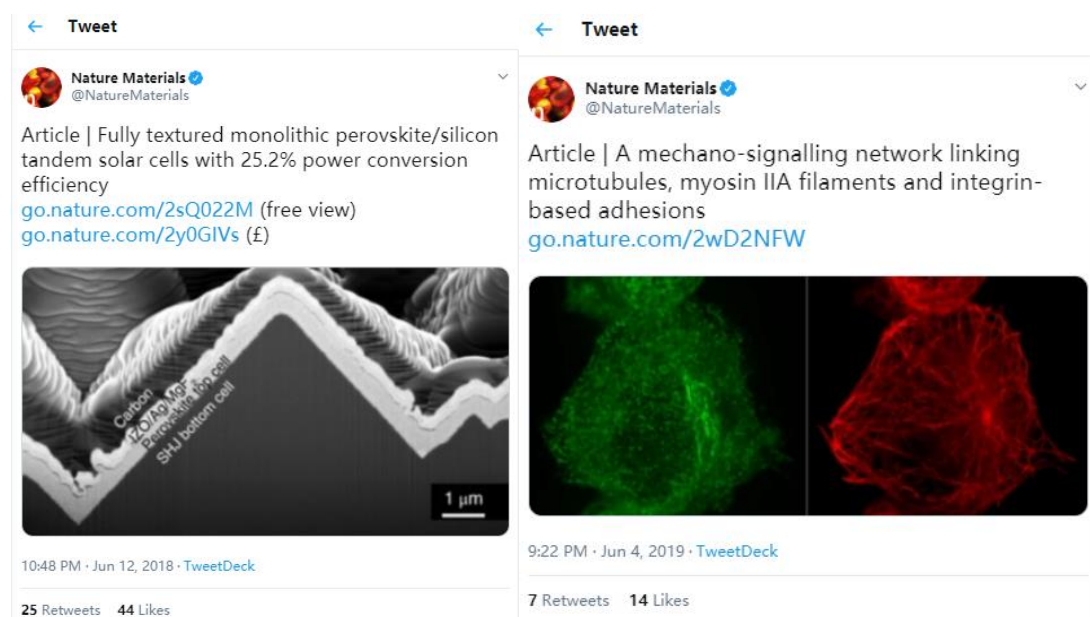

Figure 1. Screenshot of a tweet from @NatureMaterials
(Figure 1-a: two shortlinks for the same paper in one tweet pre-Oct. 2018; Figure 1-b: only one shortlink with no open access status post-Oct. 2018)

On July 23, 2019, we retrieved all the tweets and Facebook posts by @NatureMaterials from Twitter and Facebook, respectively. Only the tweets with shortlinks to one specific article were retained. The final dataset included 464 tweets, which were posted from May 4, 2017 to July 11, 2019. We then checked the Facebook posts of @NatureMaterials to find those with the same shortlinks as in the tweets and recorded the Facebook engagement data for the corresponding posts. For each OA link and corresponding paid content link, we collected Twitter engagement data and Facebook engagement data. All the Twitter and Facebook data (including the posts and engagement data) were retrieved manually.

*2.2 Data processing*

The journal sometimes posted shortlinks to the same paper repeatedly. To avoid such duplicate posts distorting the results, we discarded all papers for which there was more than one post and excluded these data to ensure that each post and each shortlink was only exposed once. There were 36 duplicate posts referring to 18 articles, so for these 18 articles, each article had been exposed twice; as the comparison to other articles would be unfair, the duplicates were dropped. We also excluded 11 abnormal posts referring to 11 articles, along with their data, because of the abnormally high number of clicks from a specific country. Thus, 47 posts were excluded from the raw dataset of 464 posts, and 417 articles were kept for analysis, including 175 articles from before Sep. 21, 2018 and 242 articles from between Sep. 21, 2018 and July 11, 2019. Each article had been exposed on both Twitter and Facebook once. Our unique research sample enabled a kind of natural matched case-control analysis.

## 3. Results

*3.1 Comparison between OA and paid content links*

We calculated and compared statistics for the three kinds of links: OA, paid content, and no status. The results are reported in Table 1 and Figure 2. The median number of clicks on OA links was 41, whereas the median number of clicks on paid content links was only 21. Therefore, OA links received about twice as many clicks as paid content links. Moreover, visitors to OA links came from a wider range of countries than those to paid content links (median: 16 vs. 11). For links not indicating OA status, the median number of clicks and of countries from which the clicks originated fell between those for OA links and paid content links.

Table 1. Comparison of different kinds of links

|  | Open access link | Paid content link | Link without status |
|---|---|---|---|
| Median number of clicks | 41 | 21 | 35 |
| Median number of countries originating clicks | 16 | 11 | 15 |

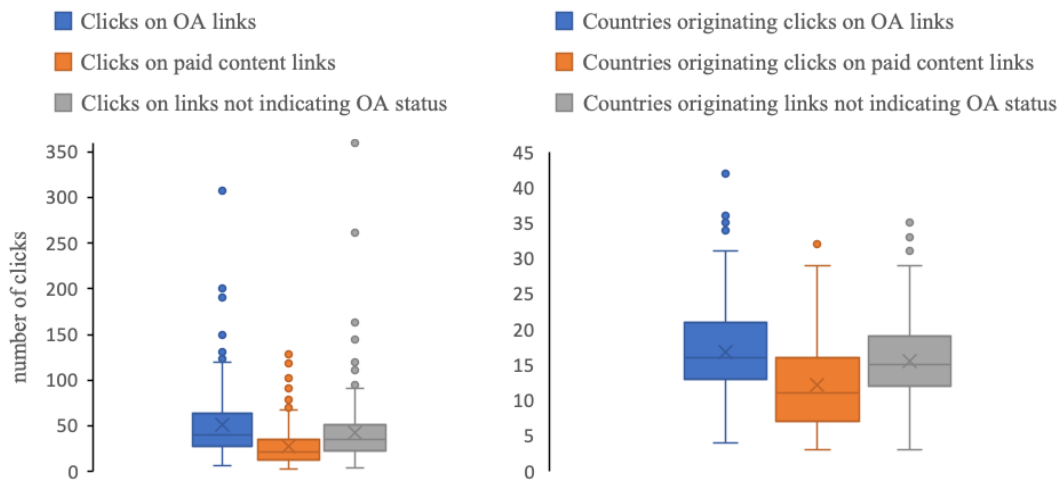

Figure 2. Comparison of open access, paid content, and no-status links

We next compared the difference in the number of clicks between the two types of links (OA and paid content) embedded in the same post and referring to the same paper, as shown in Figures 3 and 4. Each dot in the figures corresponds to a link, with blue dots indicating the OA links and orange dots representing the paid content links; the vertical axis indicates the number of clicks, and the dots are arranged along the horizontal axis by the number of OA links in descending order. OA links received more clicks and attracted visitors from more countries than paid content links for most tweets. However, for a few tweets, paid content links performed better than OA links.

To conduct a further paired difference test and verify whether OA links received more clicks than paid content links, we tested the differences in the matched-pair data. As the result reported a p-value of $< .01$, we can reject the null hypothesis that the two kinds of links would attract an equal number of clicks, meaning that the number of clicks on OA links had a median significantly different from that on clicks on paid content links. Therefore, the OA links attracted more clicks than the paid content links.

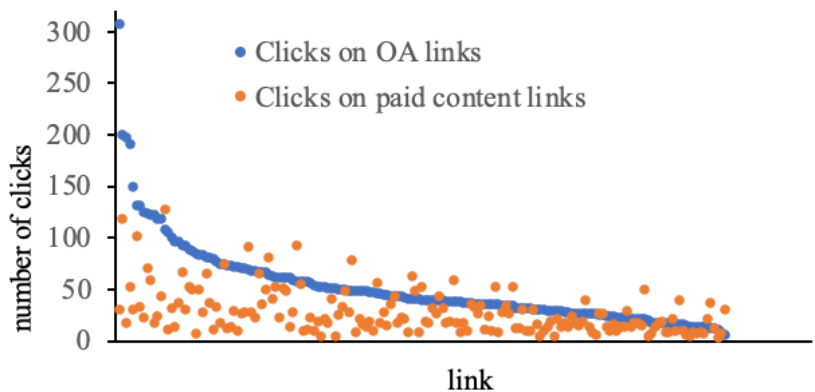

Figure 3. Comparison between clicks on open access links and clicks on paid content links embedded in the same post

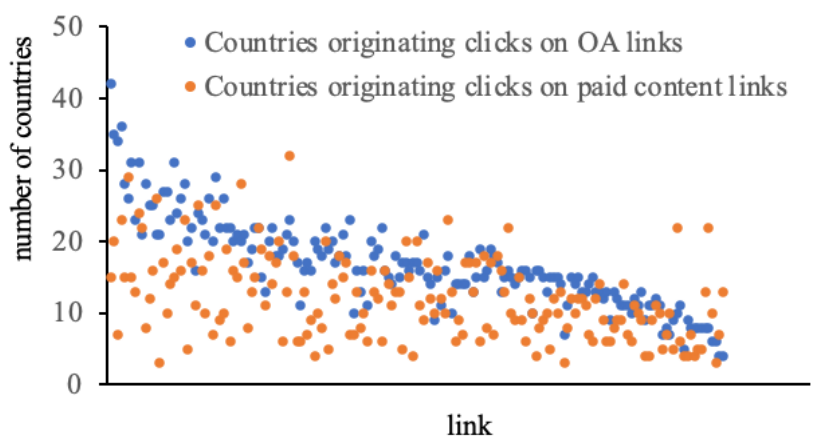

Figure 4. Comparison between the number of countries originating visits to open access links and that to paid content links embedded in the same post

We also compared clicks from the different social media platforms, as reported in Table 2 and Figure 5. On Twitter, the median number of clicks for an OA link was 24, while the median number of clicks for a paid content link was only 5. On Facebook, however, the median number of clicks for an OA link was only 7.5, whereas the median number of clicks for a paid content link was 9, which is an interesting phenomenon. The difference in performance between Twitter and Facebook may be due to the differences between the two platforms. We suppose that most of the followers of scholarly journals are researchers; when they share on Facebook, they are connecting with friends (colleagues), which could mean that they are able to target people who have access to paid content and ignore open access links. However, when Twitter users post, they are (more often than on Facebook) connecting with strangers (the general public), who usually do not have access to subscription-based scholarly resources and are interested in OA articles.

Table 2. Comparison of different kinds of links

|  | Open access links | | Paid content links | |
| --- | --- | --- | --- | --- |
| Type | Twitter | Facebook | Twitter | Facebook |
| Median number of clicks | 24 | 7.5 | 5 | 9 |
| Median number of countries originating visits | 16 | | 12 | |

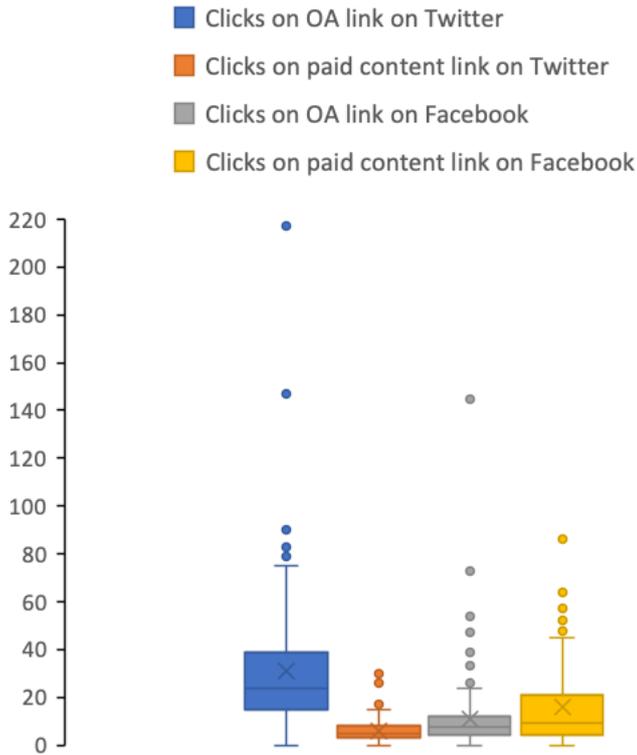

Figure 5. Comparison of clicks between links shared on Twitter and Facebook

*3.2 Regression analysis*

To explore whether OA status promotes visits to articles through social media channels, we employed regression analysis to examine the relationship between link type and the number of clicks on links. Each post included two links: one with OA status and the other with paid-content status. Therefore, we split the sample into these two groups. The number of clicks was considered the dependent variable, while social media engagements, including retweets and likes, and are the independent variables. Considering that the number of link clicks is a non-negative integer, count data models are better than linear models. Common count data models include Poisson regression, negative binomial regression, and the zero-inflated model. Because the dataset does not have a large number of zero values, it is inappropriate to employ the zero-inflated model. Therefore, Poisson regression and negative binomial regression have been adopted; the regression results are shown in Tables 3 and 4. The LR (likelihood ratio) test in the last column indicates the over-dispersion problem, meaning that we should use negative binomial (NB) regression. Moreover, NB regression has a greater Log-likelihood value and lower AIC and BIC values than Poison regression, which also suggests that NB regression is more appropriate than the other approaches. Therefore, we used the following negative binomial regression model:

$$E(Y_i) = \exp(b_0 + b_1 R + b_2 L)$$

Where Y indicates the number of clicks, R refers to retweets or Facebook shares, L is likes, and $b_1$ and $b_2$ are the regression coefficients. Stata 15.0 was employed in the analysis. The regression results are shown in Tables 3 and 4.

Table 3. Regression analysis results for Twitter

| Type | Variables | Poisson | NB |
|---|---|---|---|
| Open access link | Retweets | 0.0326*** (0.009) | 0.0449*** (0.001) |
| | Likes | 0.0356*** (0.000) | 0.0415*** (0.000) |
| | Cons. | 2.8164*** (0.000) | 2.6988*** (0.000) |
| | LR test | | 0.2029 [0.1305,0.2753] |
| | Log-likelihood | -1002.927 | -686.8115 |
| | AIC | 11.4849 | 7.8721 |
| | BIC | 11.521 | 7.9083 |
| Paid content link | Retweets | 0.0413** (0.035) | 0.0482** (0.030) |
| | Likes | 0.0103 (0.411) | 0.0107 (0.451) |
| | Cons. | 1.4692*** (0.000) | 1.4236*** (0.000) |
| | LR test | | 0.4207 [0.2570,0.5844] |
| | Log-likelihood | -591.3516 | -481.0483 |
| | AIC | 6.7812 | 5.520552 |
| | BIC | 6.8173 | 5.556720983 |

Table 4. Regression analysis results for Facebook

| Type | Variables | Poisson | NB |
|---|---|---|---|
| Open access link | Shares | -0.0021(0.916) | 0.0322* (0.098) |
| | Likes | 0.0782***(0.000) | 0.0633*** (0.000) |
| | Cons. | 1.6324***(0.000) | 1.635*** (0.000) |
| | LR test | | 0.5083 [0.3593,0.6573] |
| | Log-likelihood | -859.5013 | -536.1919 |
|  | AIC | 9.8457 | 6.1508 |
|  | BIC | 9.8819 | 6.1869 |
| Paid content link | Shares | 0.1176*** | 0.1481*** |
|  |  | (0.000) | (0.000) |
|  | Likes | 0.0209 | 0.0342* |
|  |  | (0.133) | (0.075) |
|  | Cons. | 2.0751*** | 1.8048*** |
|  |  | (0.000) | (0.000) |
|  | LR test |  | 0.7265 |
|  |  |  | [0.51039,0.9427] |
|  | Log-likelihood | -1223.8578 | -604.7439 |
|  | AIC | 14.0098 | 6.9342 |
|  | BIC | 14.0460 | 6.9704 |

Focusing first on Twitter, the results show that both retweets and likes have a positive relationship with the number of clicks on links. For OA links, the regression coefficients of retweets and likes are 0.0449 and 0.0415, respectively. However, for paid content links, likes are not statistically significant. The results indicate that a tweet being liked is related to more clicks on OA links than on paid content links for the same paper. This confirms the results of the descriptive analysis, demonstrating the OA effect based on matched case-control analysis. On Facebook, both shares and likes positively affected the number of clicks for both OA links and paid content links. As explained earlier, this may be attributable to the differences between the two platforms.

*3.3 Country analysis*
In Table 5, we list the top 30 countries/regions originating visits to all paywalled links or free links embedded in the same post. For clicks on paid content links, the top three countries were the US, Germany, and the UK; by contrast, for clicks on OA links, Japan surpassed Germany and ranked third. South Korea ranked fourth for paid content links and seventh for OA links. Meanwhile, China did not rank highly for either paid content or OA links, which can be explained by the unavailability of both Twitter and Facebook in mainland China.

Comparing OA links and paid content links, the number of clicks originating from all countries/regions for OA links was greater than that for paid content links. For example, for visitors based in the US, the numbers of clicks on paid content links and OA links were 1,086 and 2,330, respectively. Similarly, OA links received 2.47 times as many clicks as paid content links in the UK, and the same ratio in Japan was 1.97. For the top 30 countries/regions, the median value of the ratio was 1.66. We also calculated the ratio of clicks on OA to paid content links for the top 30 countries. China ranked first, followed by Russia, Switzerland, Spain, and Poland.

Table 5. Top 30 countries/regions originating visits to paywalled or free links

| Rank | Country/Region | Clicks on paid content links | Country/Region | Clicks on OA links | Country/Region | Ratio on OA/paid content |
|---|---|---|---|---|---|---|
| 1 | US | 1,086 | US | 2,330 | China | 5.00 |
| 2 | Germany | 374 | UK | 925 | Russia | 4.83 |
| 3 | UK | 313 | Japan | 617 | Switzerland | 3.65 |
| 4 | South Korea | 285 | Germany | 573 | Spain | 3.38 |
| 5 | India | 217 | France | 425 | Poland | 3.24 |
| 6 | Japan | 209 | India | 343 | UK | 2.96 |
| 7 | France | 187 | South Korea | 320 | Japan | 2.95 |
| 8 | Canada | 185 | Canada | 310 | France | 2.27 |
| 9 | Taiwan | 174 | Switzerland | 252 | US | 2.15 |
| 10 | Italy | 128 | Spain | 240 | Sweden | 2.04 |
| 11 | Turkey | 121 | Italy | 179 | Australia | 1.84 |
| 12 | Singapore | 91 | Turkey | 179 | Netherlands | 1.80 |
| 13 | Brazil | 83 | Australia | 147 | Belgium | 1.71 |
| 14 | Australia | 80 | Singapore | 146 | Canada | 1.68 |
| 15 | Thailand | 76 | Taiwan | 143 | Singapore | 1.60 |
| 16 | Netherlands | 74 | Netherlands | 133 | India | 1.58 |
| 17 | Spain | 71 | China | 115 | Germany | 1.53 |
| 18 | Switzerland | 69 | Brazil | 111 | Mexico | 1.49 |
| 19 | Philippines | 60 | Sweden | 92 | Turkey | 1.48 |
| 20 | Hong Kong | 51 | Thailand | 87 | Austria | 1.45 |
| 21 | Sweden | 45 | Philippines | 68 | Ireland | 1.42 |
| 22 | Vietnam | 45 | Hong Kong | 64 | Italy | 1.40 |
| 23 | Austria | 42 | Austria | 61 | Brazil | 1.34 |
| 24 | Israel | 42 | Mexico | 58 | Hong Kong | 1.25 |
| 25 | Greece | 41 | Russia | 58 | Israel | 1.19 |
| 26 | Mexico | 39 | Argentina | 53 | Thailand | 1.14 |
| 27 | Russia | 38 | Belgium | 53 | South Korea | 1.12 |
| 28 | Ireland | 33 | Israel | 50 | Greece | 1.05 |
| 29 | Belgium | 31 | Finland | 48 | Vietnam | 0.93 |
| 30 | Sri Lanka | 26 | Ireland | 47 | Taiwan | 0.82 |

To estimate the promotion effect of OA in different kinds of countries/regions, we classified them into two categories following the criteria developed by the World Economic Outlook (Nam, 2018), i.e. developed countries and developing countries. As reported in Table 5, for both categories, OA links received far more clicks than paid content links: the ratio was 1.92 in developed countries and 1.60 in developing countries. The further paired difference test for both developed and developing countries reported a p-value of < .05. Thus, OA links attracted more clicks than paid content links in both developed and developing countries (as shown in Table 6).

For paid content links, clicks from developed countries accounted for 83.33% of all clicks, whereas clicks from developing countries accounted for 16.67%. For OA links, the percentage of clicks from developed countries rose to 85.52%, and the percentage of developing countries fell to 14.48%. The paired difference test in Table 7 suggests that developed countries generate more clicks on both paid content links and OA links than developing countries, as shown in Figure 6. OA evidently promotes clicks in both the developed and developing world, but the beneficial impact appears to be slightly greater for developed countries.

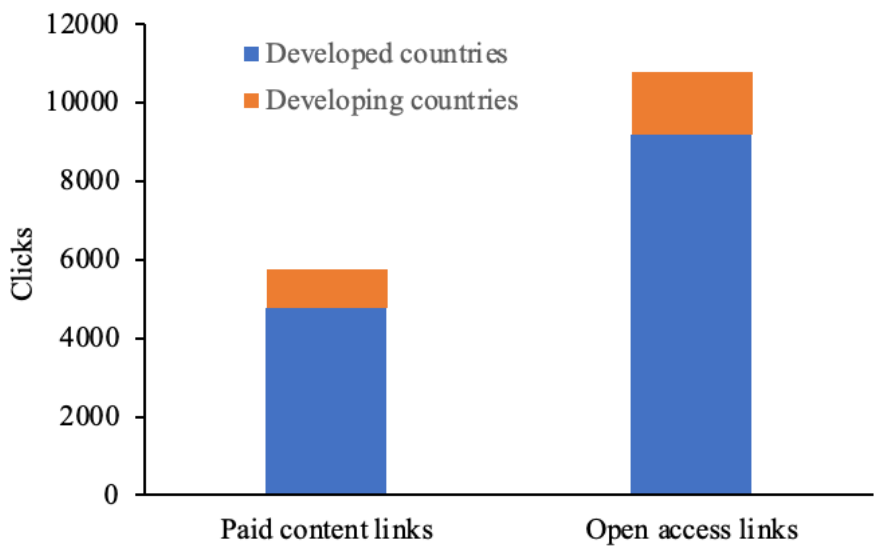

Figure 6. Clicks from developed and developing countries on paid content and OA links

Table 6. Clicks from developed and developing countries on paid content and OA links

|  | Paid content links (%) | Open access links (%) |
| --- | --- | --- |
| Developed countries | 4,030 (83.33%) | 7,729 (85.52%) |
| Developing countries | 806 (16.67%) | 1,309 (14.48%) |

Table 7. Paired difference test of clicks from developed and developing countries on paid content and OA links

|  | Open access link | Paid content link |  |
| --- | --- | --- | --- |
| Developed countries | 123.90±42.83 | 63.67±19.99 | p = .0138* |
| Developing countries | 30.28±9.31 | 18.77±5.96 | p = .0099** |
|  | p = .0364* | p = .0347* |  |

** $p < .01$, * $p < .05$

## 4. Discussion and Conclusion

This study finds that OA links tend to generate more article visits than paid content links on social media platforms. On average, OA links receive twice as many clicks as paid content links. In particular, when the post does not indicate that a linked article is OA, there is an obvious drop in performance against the clicks on links indicating the OA status. As confirmed in many previous studies, OA articles attract more downloads, social media attention, and citations than non-OA articles; we find that the OA effect also applies for links to scholarly articles. OA links receive more clicks than paid content links, which means the articles attract more readers.

Previous studies of social media sharing of scholarly articles noted that "only a few tweets reflected interest for the content of a paper" (Tahamtan & Lutz Bornmann, 2020) and suggested that social media engagement does not mean anything because people tweet/retweet whatever they see without paying attention to the content of the tweet (Robinson-García et al., 2017), let alone clicking the links or reading the paper. However, in this study, we observed a positive relationship between social media engagement and clicks on research articles originating in social media referrals. Moreover, the situations differ between Facebook and Twitter. For Facebook, social media reactions positively affect clicks on both OA links and paid content links. For Twitter, by contrast, only retweeting affects clicks on paid content links.

OA links attract visitors coming from a wider range of countries than paid content links. OA benefits all countries, but its positive impact is slightly greater for developed than developing countries. We find that OA links obtain more clicks than paid content links in both developed and developing countries, but the difference between clicks was greater in the former. This can be explained by country research capacity: developed countries often have more scientists and R&D personnel and thus generate more accesses for OA papers than developing countries. In the social media context, people face many restrictions and may find it inconvenient to access research papers behind paywalls, no matter whether they are from developed or developing countries, so the OA version is a better and more convenient choice. People in developed countries are not indifferent to having to pay to read research papers. Therefore, with the greater research capacity along with the larger number of scientists, the demand for the OA version of research papers is greater in developed than in developing countries.

Combining social media with OA seems to boost the diffusion of scholarly work. Social media exposure of scholarly articles may promote the use of research outputs, as people click links to scholarly articles embedded in social media posts. We find a significant correlation between social media engagement (e.g., retweets, likes) and clicks. However, OA links drive higher numbers of clicks than links to paid content. The subscription fees of academic journals (which can reach hundreds of thousands of dollars per year) are too high for many academic institutions in poor countries, making it difficult for scientists affiliated with them to access paid content. Moreover, scientists often need to use a VPN client to access licensed resources while they are off campus, and having to switch to a VPN connection when clicking on a link to a subscription article through Twitter or Facebook may affect the user experience. From a public perspective, science should be open to the whole of society: public disclosure may promote citizen engagement and even active participation in scientific experiments and data collection. Social media offers a direct way for scientists to connect with the general public. Although scientists are increasingly embracing social media in their professional lives, the majority of those who engage with scientific content on social media belong to the general public. The exposure of scholarly articles on social media pushes scientists' research outputs to the general public directly; however, when links to paid content are shared, members of the general public are unable to access the articles by clicking on the links provided.

Open access seeks to return science to its original purpose: to help advance and improve society. Open access is important to a number of groups, including researchers, students, practitioners, the public, policymakers, and developing countries. By providing immediate and unrestricted access to the latest research findings, researchers can keep up to date with developments in their field more effectively and accelerate the pace of discovery. Open access accelerates not only research but the translation of research into benefits for the public by sharing results with practitioners who can apply the new knowledge. Open access can provide a wealth of high quality, freely available material useful to students for eLearning.

The usual way people share a free version of a paper is through preprints. If preprints are accepted for publication, authors may link to their formal publication from the preprint via its Digital Object Identifier (DOI). Undoubtedly, the peer-reviewed formal publication will be better than its preprint, as the preprints may have differences from the final formal publication, sometimes even including mistakes or errors not corrected in the preprints. However, the formal publication may be behind a paywall when people share a link to the article. It would be very useful if people could share and let people access the formal publications behind the paywall easily and legally. There are some initiatives from publishers trying to solve this problem. For example, JSTOR's free read-only service, which is designed primarily for people who are unaffiliated with an institution, offers a way for independent researchers to read articles from JSTOR. Anyone with a free personal JSTOR account can read up to six journal articles online every 30 days. The authors of articles in certain Wiley journals can share a 'read-only' online version of their article (authenticated Wiley subscribers will get full access). Taylor & Francis Online is currently conducting a trial of an eReader to enable people to share free read-only access to journal articles with friends and colleagues; a trial program including 22 journals began in January 2019. As a free content-sharing initiative supported by ReadCube and provided by Springer Nature, SharedIt enables people to post links to free-to-read versions of subscription research articles anywhere, including social media platforms, repositories, and websites. For those audiences without access to subscription articles, these services provide a compromise that can meet the requirements of free-to-read. In the social media ecosystem, free-to-read anywhere and anytime is the most important aim.

There are some limitations to this study. First, only *Nature Materials* has this kind of matched-pair data, so the sample size in this study is relatively small. Second, we cannot reach a conclusion about the causal relationships between social media exposure and number of clicks. Third, why are Facebook users insensitive to paid content? We could not determine the exact reason based on the data, so we answered the question based on our best guess. Finally, the effect of OA articles may be different from the effect of OA links (links to read-only versions of full-text subscription articles); this is an issue that needs further exploration.

**Acknowledgments**
This research was supported by the "National Natural Science Foundation of China" (71673038, 71974029) and the project of ''Fundamental Research Funds for the Central Universities'' (300102119301).

**References**
Allen, H. G., Stanton, T. R., Di Pietro, F., & Moseley, G. L. (2013). Social media release increases dissemination of original articles in the clinical pain sciences. *PloS one*, *8*(7), e68914.

Björk, B. C., Laakso, M., Welling, P., & Paetau, P. (2014). Anatomy of green open access. *Journal of the Association for Information Science and Technology*, *65*(2), 237-250.

Bornmann, L. (2014). Validity of altmetrics data for measuring societal impact: A study using data from Altmetric and F1000Prime. *Journal of Informetrics*, *8*(4), 935-950.

Cui, Y., Wang, X., Xu, S., Hu, Z. & Zhang, C. (2018). Evaluating the influence of social media exposure of scholarly articles: Perspectives of social media engagement and click metrics, STI 2018: 23rd International Conference on Science and Technology Indicators, Leiden.

Darling, E.S., Shiffman, D., Côté, I., & Drew, J.A. (2013). The role of Twitter in the life cycle of a scientific publication. *Ideas in Ecology and Evolution*, 6, 32–43. http://doi.org/10.7287/peerj.preprints.16v1

Ezema, I. J., & Onyancha, O. B. (2017). Citation impact of health and medical journals in Africa: does open accessibility matter?. *The Electronic Library*, *35*(5), 934-952.

Harnad, S, Brody, T, Vallières, F, Carr, L, Hitchcock, S, Gingras, Y, … Hilf, E (2008). The Access/Impact Problem and the Green and Gold Roads to Open Access: An Update. Serials Review, 34(1), 36–40. doi:10.1016/j.serrev.2007.12.005.

Haustein, S., Sugimoto, C. and Larivière, V. (2015), Guest editorial: social media in scholarly communication, Aslib Journal of Information Management, 67 (3). doi:10.1108/AJIM-03-2015-0047

Hawkins, C.M., Hillman, B.J., Carlos, R.C., Rawson, J.V., Haines, R., & Duszak, R. (2014). The impact of social media on readership of a peer-reviewed medical journal. *Journal of the American College of Radiology, 11 11*, 1038-43.

Hawkins, C.M., Hunter, M., Kolenic, G.E., & Carlos, R.C. (2017). Social Media and Peer-Reviewed Medical Journal Readership: A Randomized Prospective Controlled Trial. *Journal of the American College of Radiology, 14 5*, 596-602 .

Kelly, B., & Delasalle, J. (2012). Can LinkedIn and Academia. edu enhance access to open repositories? In *OR2012: The Seventh International Conference on Open Repositories* . Edinburgh, Scotland: University of Bath. Retrieved from http://opus.bath.ac.uk/30227

Koler-Povh, T., Južnič, P., & Turk, G. (2014). Impact of open access on citation of scholarly publications in the field of civil engineering. *Scientometrics*, *98*(2), 1033-1045.

Laakso, M, Welling, P, Bukvova, H, Nyman, L, Björk, B, & Hedlund, T (2011). The Development of Open Access Journal Publishing from 1993 to 2009. PLoS ONE, 6(6), e20961. doi:10.1371/journal.pone.0020961.

Lévy, J. (2018). Social media for scientists. *Nature Cell Biology, 20*, 1329.

Nam, C. W. (2018). World Economic Outlook for 2018 and 2019. In *CESifo Forum* (Vol. 19, No. 1, pp. 51-51). Institut für Wirtschaftsforschung (Ifo).

O'Kelly, F., Nason, G.J., Manecksha, R.P., Cascio, S., Quinn, F.M., Leonard, M.P., Koyle, M.A., Farhat, W., & Leveridge, M.J. (2017). The effect of social media (#SoMe) on journal impact factor and parental awareness in paediatric urology. *Journal of pediatric urology, 13 5*, 513.e1-513.e7.

Ottaviani, J. (2016). The Post-Embargo Open Access Citation Advantage: It Exists (Probably), It's Modest (Usually), and the Rich Get Richer (of Course). *PLoS ONE,* 11(8): e0159614. https://doi.org/10.1371/journal.pone.0159614

Priem, J., Taraborelli, D., Groth, P., & Neylon, C. (2010). Altmetrics: A manifesto. 26 October 2010. http://altmetrics.org/manifesto

Rainie, L., Funk, C., & Anderson, M. (2015). How scientists engage the public. *Pew Research Center*.

Robinson-Garcia, N., Costas, R., Isett, K.R., Melkers, J., & Hicks, D. (2017). The unbearable emptiness of tweeting—About journal articles. *PLoS ONE,* e0183551. https://doi.org/10.1371/journal.pone.0183551.

Schöpfel, J. (2017). Open Access to Scientific Information in Emerging Countries. *D-Lib Magazine*, *23*(3), 5.

Sotudeh, H., & Estakhr, Z. (2018). Sustainability of open access citation advantage: the case of Elsevier's author-pays hybrid open access journals. *Scientometrics*, *115*(1), 563-576.

Suber, P., & Arunachalam, S. (2005). Open access to science in the developing world. *World-Information City*, October 17, 2005. http://legacy.earlham.edu/~peters/writing/wsis2.htm

Sugimoto, C. R., Work, S., Larivière, V., & Haustein, S. (2017). *Scholarly use of social media and altmetrics: A review of the literature. Journal of the Association for Information Science and Technology, 68(9), 2037–2062.* doi:10.1002/asi.23833

Tahamtan, I., & Bornmann, L. (2020). Altmetrics and societal impact measurements: Match or mismatch? A literature review. *Profesional De La Informacion, 29*(1), e290102. https://doi.org/10.3145/epi.2020.ene.02

Tennant, J. P, Waldner, F, Jacques, D. C, Masuzzo, P, Collister, L. B, & Hartgerink, C. H. J (2016). The academic, economic and societal impacts of Open Access: an evidence-based review. F1000 Research, 5, 632. doi:10.12688/f1000research.8460.2.

Trueger, N.S. (2018). Medical Journals in the Age of Ubiquitous Social Media. *Journal of the American College of Radiology, 15 1 Part B*, 173-176.

Tsou, A., Bowman, T. D., Ghazinejad, A., & Sugimoto, C. R. (2015, June). Who tweets about science?. *ISSI Conference*.

Wang, X., Liu, C., Mao, W., & Fang, Z. (2015). The open access advantage considering citation, article usage and social media attention. *Scientometrics*, *103*(2), 555-564.

Wang, X., Cui, Y., Xu, S., & Hu, Z. (2018). The state and evolution of Gold open access: A country and discipline level analysis. *Aslib Journal of Information Management*, *70*(5), 573-584.

Wang, X., Fang, Z., & Guo, X. (2016). Tracking the digital footprints to scholarly articles from social media. *Scientometrics*, *109*(2), 1365-1376.

Wang, X., Cui, Y., Li, Q., & Guo, X. (2017). Social media attention increases article visits: An investigation on article-level referral data of PeerJ. *Frontiers in Research Metrics and Analytics*, *2*, 11.https://doi.org/10.3389/frma.2017.00011

Zedda, M., & Barbaro, A. (2015). Adoption of web 2.0 tools among STM publishers. How social are scientific journals? Journal of the EAHIL, 11, 9–12.